\documentclass[english]{spieman}
\usepackage{amsmath}
\usepackage{mathptmx}
\usepackage[T1]{fontenc}
\usepackage[latin9]{inputenc}
\usepackage{geometry}
\usepackage{color}
\definecolor{note_fontcolor}{rgb}{0.800781, 0.800781, 0.800781}
\usepackage{graphicx}
\usepackage{subscript}

\makeatletter

\@ifundefined{date}{}{\date{}}
\makeatother

\usepackage{babel}
\begin{document}

\title{\centering Computing with Dynamical Systems Based on 
Insulator-Metal-Transition Oscillators}

\author[a]{Abhinav Parihar}
\author[b]{Nikhil Shukla}
\author[b]{Matthew Jerry}
\author[b]{Suman Datta}
\author[a]{Arijit Raychowdhury}
\affil[a]{Georgia Institute of technology}
\affil[b]{University of Norte Dame}

\maketitle

\begin{abstract}
In this paper we review recent work on novel computing paradigms using coupled oscillatory dynamical systems. We explore systems of relaxation oscillators based on linear state transitioning
devices, which switch between two discrete states with hysteresis. By harnessing the dynamics of complex, connected systems we embrace the philosophy of ``let physics do the computing'' and demonstrate how complex phase and frequency dynamics of such systems can be controlled, programmed and observed to solve computationally hard problems. Although our discussion in this paper is limited to Insulator-to-Metallic (IMT) state transition devices, the general philosophy of such computing paradigms can be translated to other mediums including optical systems. We present the necessary mathematical treatments necessary to understand the time evolution of these systems and demonstrate through recent experimental results the potential of such computational primitives.

\end{abstract}

\section{Introduction}

Computing is the backbone of the modern society; from economics to security, scientific advancement to social welfare, each and every aspect of our lives has been enriched by the technology revolution. We have enjoyed the benefits of Moore's Law over the last four decades as technology scaling brought the power of supercomputers to our smartphones. With increasing challenges in scaling, came ground-breaking innovations in the transistor technology. As we look ahead, limits of traditional scaling are already in sight. The demise of Dennard scaling and slowing down of Moore's Law have further exposed the fundamental scaling limitations of the Von Neumann execution models of computing. This transition is accompanied by the realization that in a fast evolving, socially interconnected world, we are observing a seismic shift in the amount of unstructured data that need to be processed in real-time; and consequently future systems will be limited by the energy growth of data movement rather than compute. Therefore, we need fundamentally new approaches to sustain the exponential growth in performance beyond the end of the CMOS roadmap. This will require new execution models coupled with new devices to implement them. In particular, we observe that new models that deal with data analytics have compute and storage interleaved in a fine grained manner - not separated as in the Von Neumann world. Moving forward, computing technology will heavily penalize separation of data and compute and we need to marry them in better ways to handle emergent applications. This may necessitate that barriers of abstraction are broken. Next generation of computational models should map natively to the physics and dynamics of the physical devices, without a Boolean abstraction. Of course, it is a fool's endeavor to assume that the Von Neumann architecture will perish; rather newer computing models and dedicated hardware accelerators will supplement traditional Von Neumann machines in ``data centric'' tasks.

In today's computing landscape, ever harder problems are being encountered each day. From social networks to graph analytics, from weather prediction to scientific computation, computationally hard problems are present everywhere. Some of these problems are challenging just because of the sheer
size of the data-sets. But many other
problems like optimization problems, though small in size, are intractable
because of their inherent complexity and often combinatorial
nature of solution spaces\cite{garey1979computers}. The sequential
Von Neumann machine has been useful for most problems
we have encountered in the past, but as the complexity of problems increase, significant research is underway for
alternative paradigms, architectures and hardware which can be used
for solving complex problems more efficiently. Most of these efforts
borrow essential ideas from natural computing processes. These include
distributed computing, distributed memory, integrated processing and
memory, shifting information representation from symbolic to physically
meaningful quantities, and switching from sequential discrete time
to continuous time dynamics.

Among the different problems of interest, associative computing, scientific computing (including solution of coupled Partial Differential Equations (PDEs)) and optimizations form important classes of data-centric as well as model-centric computations. Active research in all these areas suggest that analog or continuous time systems may offer alternative, faster and more energy-efficient solutions that their traditional digital counterparts. For example, in case of hard optimization problems
alternative paradigms and architectures include, but are not limited
to, cellular automata\cite{wolfram1986theoryand}, quantum computing\cite{shor1997polynomialtime},
Ising model based systems\cite{lucas2014isingformulations,wang2013coherent},
neural networks\cite{hopfield1985textquotedblleftneuraltextquotedblright},
stochastic searching architectures\cite{mostafa2015aneventbased}
and memcomputing\cite{traversa2015memcomputing}. Among architectures
for solving PDEs cellular neural networks (CNN) have been studied\cite{chua1988cellular2,chua1988cellular}.
Some studies have also suggested using cellular automata for solving
PDEs\cite{toffoli1984cellular}. 

The basic philosophy of most of these networks for optimization, e.g.
artificial neural networks, is to first come up with an energy function which
can be a penalty function or a rewarding function depending on how
far the current solution is from the optimal solution\cite{hopfield1985textquotedblleftneuraltextquotedblright}.
The next step is to tune the parameters of the network such that as
time evolves the dynamics of the system decreases the penalty function,
or increases the reward. But even if such a massively parallel system
is devised which can solve a NP-hard optimization problems, exponential
resources of space, time or precision will be required\cite{siegelmann1995computation,vergis1986thecomplexity}.
Another direction is often explored where instead of trying to solve
the optimization exactly, an approximation is targeted which works
well for \textit{most} problems on average and allows less optimal
solutions in harder instances. There can be other kinds of trade-offs like
in the case of Hopfield networks where even though the optimal solution
maps exactly to global minima, there can be too many local minima
where the system can get trapped. Finally, the physical layer of computing, namely the semiconductor device platform needs to be able to support such systems and the CMOS transistor is not always an optmial device choice.

Recently, the development of novel phase transition materials like
Vanadium Dioxide (VO\textsubscript{2}) and corresponding electronic devices, which show insulator-to-metal
(IMT) transitions\cite{imada1998metalinsulator} have sparked interest
in creating compact relaxation oscillators\cite{shukla2014synchronized,shukla2014pairwise}.
These oscillators, when coupled to each other exhibit phase synchronization which
can be used for phase based computing. Such new kinds of devices present
interesting opportunities to create systems with novel synchornization dynamics. The
impact of using such devices as basic units in circuits can break the abstraction between the physical and the algorithmic layers of computing. It should be noted that the synchronization dynamics of coupled oscillators not only have a wide variety of applications in engineering but they also explain many natural, chemical and biological synchronization phenomena like the synchronized flashing of fireflies, pacemaker cells in the human heart, chemical oscillations, neural oscillations, and laser arrays, to name a few. These novel computing primitives, of course, are neither  drop-in replacements for CMOS transistors nor straightforward extensions of the existing computing architectures. It requires rethinking of the basic computational entities in new kinds
of system architectures. In this paper we review some computational models
using coupled relaxation oscillators based on VO\textsubscript{2}metal-insulator-transition
devices focusing on how the system dynamics can be modeled and the applications they can enable. We limit our discussion on applications in image processing for the sake of brevity. However, the potential of dynamical systems extend far beyond image analytics and promises to be a competing computational model for post-CMOS technologies.

\section{A Perspective on Coupled Oscillatory Networks}

The area of coupled oscillators has been dominated mostly by theoretical
models and numerical simulations, but very few successful physical
implementations. The reason being the assumptions made for analytical
simplification in those theoretical models are too difficult to realize
in practice. Also an important limitation of such systems, which is
true for any dynamical system, is that if a dynamical system is able
to solve computationally hard (NP-class) problems exactly, then it necessarily have to be
chaotic in nature which would require exponential precision in both simulations and physical implementations.

The most popular coupled oscillator models in this area are the \textit{Kuramoto}
oscillator models\cite{strogatz2000fromkuramoto} which rely on sinusoidal
oscillators coupled using \textquotedblleft weak\textquotedblright{}
and linear phase coupling. A \textit{Kuramoto} system of N oscillators is described
by

\[
\dot{\theta_{i}}=\dot{\omega_{i}}+\frac{K}{N}\sum_{j=1}^{N}\sin(\theta_{j}-\theta_{i}),\,\,\,i=1,...,N
\]

where $\theta_{i}$ and $\omega_{i}$ are the phase and frequency
respectively of $i^{th}$ oscillator. Major challenges in this kind
of coupled oscillator model is the notion of weak coupling, i.e. $K\ll N$
and the idea that the coupling effects phase only without disturbing the frequencies. Moreover,
creating arrays of compact sinusoidal oscillators with many oscillators
coupled to each other pose serious challenges given the requirements
 Similar models of weak linear phase coupling were also explored for
Van Der Pol oscillators\cite{vanderpol1934thenonlinear} which have
an additional nonlinearity. But the implementation of such oscillators
is also non-trivial and the coupling behavior becomes too complicated to tackle
large connected networks\cite{rand1980bifurcation,storti1982dynamics,kopell1995antiphase,hirano2003existence}.

Nevertheless, the ability of coupled oscillatory systems to encode computing has long been realized. Associative computing, with applications in pattern detection and machine learning have been demonstrated in theory. Similarly oscillatory cellular neural networks have been show to possess extraordinary computing ability in solving problems as varied as template matching, PDEs and ODEs and so on. More recently, this effort has been augmented by advances in the development of compact oscillators in non-silicon technologies. One prominent effort is the use of Spin Torque Oscillators (STOs) coupled using spin diffusion currents and providing a computational platform for machine learning, spiking neural networks and others~\cite{7120163,cite-key-kroy}. However, the high current densities of STOs and the limited range of spin diffusion currents continue to pose serious challenges to technologists.

In our studies, we explore phase transition based relaxation oscillators with piecewise
linear dynamics, which means the system is described by different
linear dynamical systems in different ``states'' of the system. These
states are basically charging and discharging of a capacitive element. The coupling is also electronic and is accomplished by linear capacitors and/or resistors. 
The repeated switching between these states gives rise to oscillatory
behavior and the notion of phase. But because the switching is itself
determined by the state variables (voltage thresholds), and not explicitly by time, the
coupling dynamics and hence the overall system dynamics are often mathematically intractable and closed form approximations like the Kumamoto model are not possible. This further complicates the analysis when
the goal is to perform computation in phase space. Hence model development coupled with numerical analysis and an intuitive understanding of how complex systems evolve over time become essential tools to engineer such systems.  The relaxation oscillators we investigate,
are built using phase change devices, which are devices that switch
state between a metallic state with low resistance and and insulating
state with high resistance depending on the voltage across them. In the next section, we will describe simple mathematical constructs that can assist in analyzing these oscillators and the results obtained once they are coupled electrically.

\section{Relaxation oscillators based on phase change devices}

\subsection{State changing devices}

The state changing devices we consider are essentially linear conductances
(or resistances), but can transition between two conducting states
- insulating state with conductance $g_{di}$ and metallic state with
conductance $g_{dm}$. We assume that $g_{dm}\gg g_{di}$. They are
also called Insulator-Metal-Transition (IMT) devices. We use Vanadium
Dioxide (VO2) as the material choice. A state transition is triggered
by the voltage applied across the devices as well as the history.
When the voltage across the device increases above a higher threshold
voltage $V_{h}$ the device changes its state to metallic with conductance
$g_{dm}$ and when the voltage decreases below a lower threshold $V_{l}$
it changes to insulating state with conductance $g_{di}$. There is
hysteresis in switching, i.e. $V_{h}\neq V_{l}$, which means when
the voltage applied is between $V_{h}$ and $V_{l}$ the device retains
the last state it was in. An internal capacitance is associated with
the device which ensures gradual buildup and decaying of the voltage
(and energy) across the device. Rigorous mathematical analysis of
such oscillator configurations can be found in \cite{parihar2015synchronization}.

\subsection{Single oscillator configurations}

\begin{figure}
\begin{centering}
\includegraphics{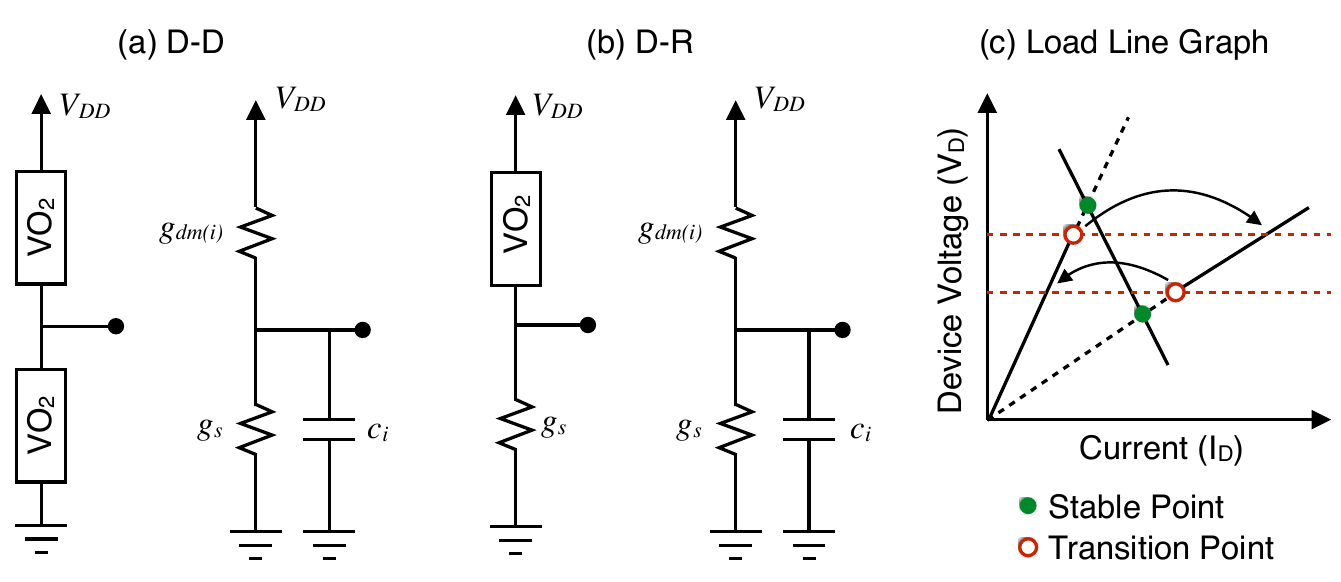}
\par\end{centering}
\caption{(a) D-D oscillator circuit configuration and its equivalent circuit
(b) D-R oscillator circuit configuration and its equivalent circuit.
(c) Load line graph for the D-R circuit and the region of operation
of D-R oscillator.\label{fig:dd_dr_load}}
\end{figure}

There can be multiple configurations/circuits of a relaxation oscillator
based on state changing device. On a simplistic level, two basic configurations
exist - (a) two state changing devices in series, called D-D where
D stands for device (figure \ref{fig:dd_dr_load}a), and (b) a state
changing device in series with a resistance, called D-R (figure \ref{fig:dd_dr_load}b).
In the former, the charging and discharging rates are equal, but they
are different in the latter. The functioning of the two circuits is
as follows.

In case of two devices in series (D-D), the two devices must be in
opposite conduction states (one metallic and the other insulating)
all the time for oscillations to occur. If the threshold voltages
$v_{l}$ and $v_{h}$ are equal for the devices and the following
condition holds:

\begin{equation}
V_{l}+V_{h}=V_{DD}\label{eq:dd-constraint}
\end{equation}

and at $t=0$ the devices are in different conduction states, then
any time one device switches, the other will make the opposite transition
as well. As $g_{dm}\gg g_{di}$, the devices can be considered as
switches which are open in insulating state and closed in metallic
with conductance $g_{dm}$. The mechanism of oscillations is essentially
charging and discharging of the internal capacitances of the devices.
The device in metallic state connects the circuit and charges (discharges)
the lumped internal capacitance. The voltage at the output node increases
(decreases) and eventually reaches the threshold voltage. Because
of (\ref{eq:dd-constraint}) both devices will switch at the same
time causing their behavior to switch. The charging (discharging)
becomes discharging (charging) and the cycle continues. The modeling
of a DD oscillator is as follows. All the lowercase voltages referred
in the paper are normalized voltages with respect to $V_{DD}$. Which
means $v_{h}=V_{h}/V_{DD}$ and $v_{l}=V_{l}/V_{DD}$. Also $v_{dd}$
is used as normalized and hence $v_{dd}=1$.

The single D-D oscillator can be described by the following set of
piecewise linear differential equations:

\[
cv'=\begin{cases}
(v_{dd}-v)g_{1dm} & charging\\
-g_{2dm} & discharging
\end{cases}
\]

where $g_{1dm}$ and $g_{2dm}$ are metallic conductances of the two
devices respectively. As $g_{di}\gg g_{dm}$ there is no term involving
$g_{di}$ in the equations. The equation can be re-written as: 
\[
cv'=-g(s)v+p(s)
\]

where $s$ denotes the conduction state of the device (0 for metallic,
and 1 for insulating) and $g(s)$ and $p(s)$ depend on the device
conduction state $s$ as follows: 
\begin{eqnarray*}
g(s) & = & \begin{cases}
g_{1dm}, & s=0\\
g_{2dm}, & s=1
\end{cases}\\
p(s) & = & \begin{cases}
g_{1dm}, & s=0\\
0, & s=1
\end{cases}
\end{eqnarray*}

For D-R oscillators, the oscillations occur due to a lack of stable
point as seen in the load line graph of figure \ref{fig:dd_dr_load}c.
Solid lines with slopes $r_{i}$ and $r_{m}$ are the regions of operation
of the device in insulating and metallic states respectively. The
system does not enter the dashed line region as a transition occurs
to the other conduction state at the red points. The stable points,
denoted by green points, are the points where the load line intersects
the I-V curve of the device. These stable points in each state lie
outside the region of operation the circuit and hence the circuit
shows self sustained oscillations. This is a much more practical configuration
from an electrical implementation point of view, as the conditions
required for oscillations are not very strict. Following similar analysis
as in the D-D oscillator case, the dynamics of the single D-R oscillator
can be described as:

\[
cv'=\begin{cases}
(v_{dd}-v)g_{dm}-vg_{s} & charging\\
-vg_{s} & discharging
\end{cases}
\]

which can be re-written as: 
\[
cv'=-g(s)v+p(s)
\]

where, 
\begin{eqnarray*}
g(s) & = & \begin{cases}
g_{dm}+g_{s}, & s=0\\
g_{s}, & s=1
\end{cases}\\
p(s) & = & \begin{cases}
g_{dm}, & s=0\\
0, & s=1
\end{cases}
\end{eqnarray*}

and $s$ denotes the conduction state of the system as before. Detailed
analysis of configurations and modeling of such oscillators can be
found in \cite{parihar2015synchronization}.

\subsection{Pairwise coupling}

\begin{figure}
\begin{centering}
\includegraphics{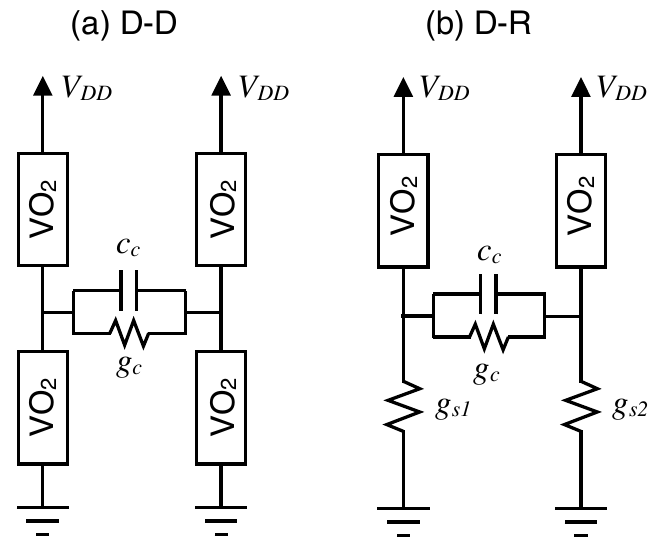}
\par\end{centering}
\caption{D-D (a) and D-R (b) coupled oscillator circuits using parallel RC
circuit as the coupling circuit.\label{fig:dd_dr_coupled}}
\end{figure}

Analysis of two coupled relaxation oscillators can give interesting
insights into how such coupling dynamics can be used in various computing
applications, and can also help understand and exploit dynamics from
complex couplings. There can be many ways in which the oscillators
can be coupled. We've looked at coupled oscillator circuits where
the oscillators are coupled through their output nodes using a capacitance,
a resistance, or a parallel RC combination (figure \ref{fig:dd_dr_coupled}).

\subsubsection{D-D oscillators}

The D-D configuration, though difficult in an electrical implementation
point of view, is very simple to analyse and gives interesting insights
about dynamics of such piecewise linear systems. Two identical D-D
oscillators coupled using a RC circuit can be modeled as follows.
When coupled, the system has 4 conduction states $s=s_{1}s_{2}\in\{00,01,10,11\}$
corresponding to the 4 combinations of $s_{1}$ and $s_{2}$. The
coupled system can be described in matrix form as: 
\begin{eqnarray*}
c_{c}Fx'(t) & = & -g_{c}A(s)x(t)+P(s)\\
x'(t) & = & -\frac{g_{c}}{c_{c}}F^{-1}A(s)\left(x(t)-A^{-1}(s)P(s)\right)
\end{eqnarray*}

where $x(t)=\left(v_{1}(t),v_{2}(t)\right)$ is the state variable
at any time instant $t$. The $2\times2$ matrices $F$ and $A(s)$,
and vector $P(s)$ are given by: 
\[
F=\left[\begin{array}{cc}
1+\alpha_{1} & -1\\
-1 & 1+\alpha_{2}
\end{array}\right]
\]
\[
\begin{array}{cc}
A(00)=\left[\begin{array}{cc}
-\beta_{11}-1 & 1\\
1 & -\beta_{21}-1
\end{array}\right], & P(00)=\left[\begin{array}{c}
\beta_{11}\\
\beta_{21}
\end{array}\right]\\
A(10)=\left[\begin{array}{cc}
-\beta_{12}-1 & 1\\
1 & -\beta_{21}-1
\end{array}\right], & P(10)=\left[\begin{array}{c}
0\\
\beta_{21}
\end{array}\right]\\
A(01)=\left[\begin{array}{cc}
-\beta_{11}-1 & 1\\
1 & -\beta_{22}-1
\end{array}\right], & P(01)=\left[\begin{array}{c}
\beta_{11}\\
0
\end{array}\right]\\
A(11)=\left[\begin{array}{cc}
-\beta_{12}-1 & 1\\
1 & -\beta_{22}-1
\end{array}\right], & P(11)=0
\end{array}
\]

Here, $\alpha_{i}=c_{i}/c_{c}$ is the ratio of the combined lumped
capacitance of $i^{th}$ oscillator to the coupling capacitance $c_{c}$,
and $\beta_{ij}=g_{ijdm}/g_{c}$ is the ratio of the metallic state
resistance of $j^{th}$ device of $i^{th}$ oscillator, where $i\in\{1,2\}$
and $j\in\{1,2\}$. The fixed point in a conduction state $s$ is
given by $p_{s}=A^{-1}(s)P(s)$ and the matrix determining the flow
(the \emph{flow matrix} or the \emph{velocity matrix}) is given by
$\frac{g_{c}}{c_{c}}F^{-1}A(s)$.

When two identical D-D oscillators are coupled using a parallel RC
circuit with coupling resistance $R_{C}$ and coupling capacitance
$C_{C}$, they can lock in-phase or anti-phase depending on the relative
values of $R_{C}$ and $C_{C}$. In the extreme case of purely capacitive
coupling with $R_{C}=0$ the anti-phase locking orbit is stable and
the in-phase locking orbit is unstable. In case of purely resistive
coupling with $C_{C}=0$ the in-phase locking orbit is stable and
the anti-phase locking orbit is unstable. For other values, the system
always have in-phase locking periodic orbit as well as anti-phase
periodic locking orbits, with the stable locking being the in-phase
locking when the coupling is close to purely resistive, and when the
coupling is close to purely capacitive the stable locking is the anti-phase
locking. Interestingly, considering the parameter space of $R_{C}$
and $C_{C}$ there exist a region with bistable orbits, i.e. both
the in-phase and anti-phase orbits are stable. In this region, the
steady state locking depends on the initial starting voltages at $t=0$
of the oscillators.

\subsubsection{D-R oscillators}

\begin{figure}
\begin{centering}
\includegraphics{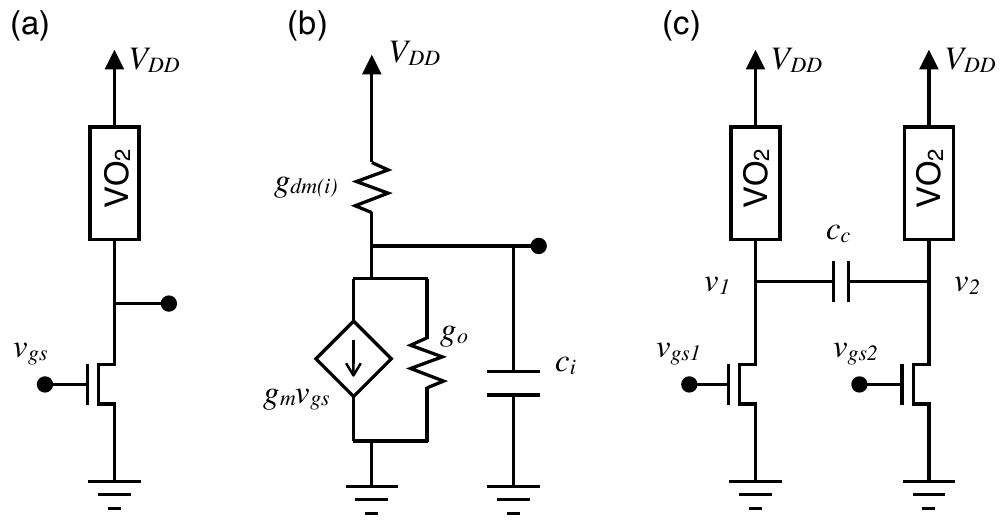}
\par\end{centering}
\caption{(a) Oscillator circuit with MOSFET in series with a VO\protect\textsubscript{2}
device (b) Small signal equivalent circuit of the oscillator with
VO\protect\textsubscript{2} in series with a MOSFET (c) Coupled oscillator
circuit with a series MOSFET in place of a series resistance.\label{fig:vo2_mosfet}}
\end{figure}

\begin{figure}
\begin{centering}
\includegraphics{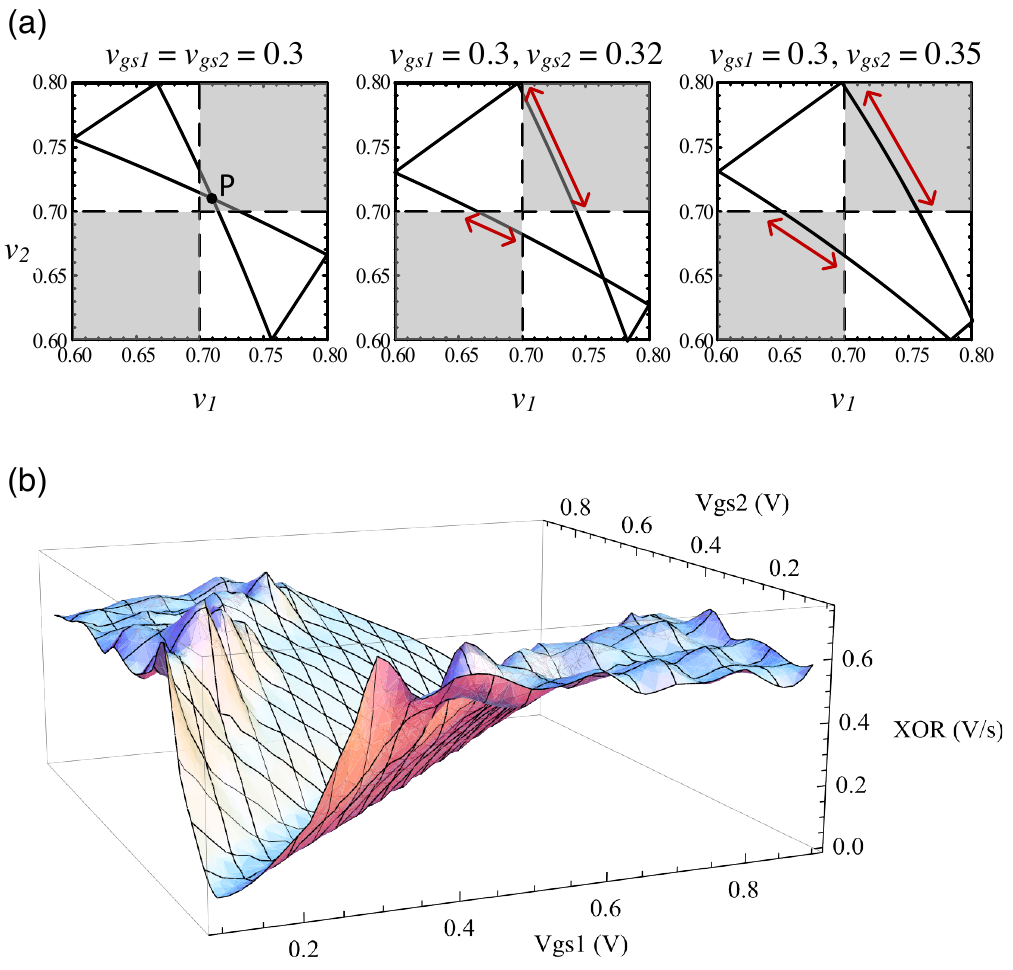}
\par\end{centering}
\caption{(a) XOR surface 3D figure\label{fig:xor_butterflies}}

\end{figure}

From a computing application point of view pairwise coupled D-R oscillators
have interesting applications. When the series resistances are replaced
by transistors as shown in figure \ref{fig:vo2_mosfet} and the coupling
is a simple capacitive coupling, a pair of coupled D-R oscillators
can be used as a analog comparator whose output has the form of a
difference norm\cite{datta2014neuroinspired,parihar2014exploiting,shukla2014pairwise}.
Typical steady state orbits of the coupled system plotted in a $v_{1}\times v_{2}$
plane are shown in figure \ref{fig:xor_butterflies}a. When $v_{gs1}-v_{gs2}$
increases, the steady state orbits of the oscillators gets deformed.
Such deformation of the steady sate orbits can be measured using a
simple averaged thresholding-and-XOR operation on the steady state
outputs of the oscillators. This averaged XOR measure is defined as
first thresholding the output to binary values, second applying XOR
operation on these binary values at every time instant and finally
averaging this XOR output over some time duration. The averaged XOR
output as a function of $v_{gs1}$ and $v_{gs2}$ is shown in figure
\ref{fig:xor_butterflies}b. The XOR surface reaches minimum value
along the line $v_{gs1}=v_{gs2}$. Within the locking range, it rises
as an even function of $v_{gs1}-v_{gs2}$ resembling a difference
norm. Outside the locking range, it averages to about 0.5. These characteristics
of the curve can be explained by realizing that the averaged XOR measure
by construction is equal to the fraction of the time the system spends
in the grey region (region where XOR output is 1\textemdash determined
by the thresholds on $v_{1}$ and $v_{2}$) in steady state as shown
in figure \ref{fig:xor_butterflies}a. It can be seen that the XOR
measure should have least value in the symmetric case $v_{gs1}=v_{gs2}$
when the system locks out-of-phase and should increase as values diverge.
Such a system can be used as an analog comparator with output as a
difference norm. Arrays of such comparators can be used for template
matching applications where element-wise comparisons suffice to decide
a match.

\section{IMT Oscillators}

\subsection{IMT Basics}
\begin{figure}[t]
\begin{centering}
\includegraphics[width=\textwidth]{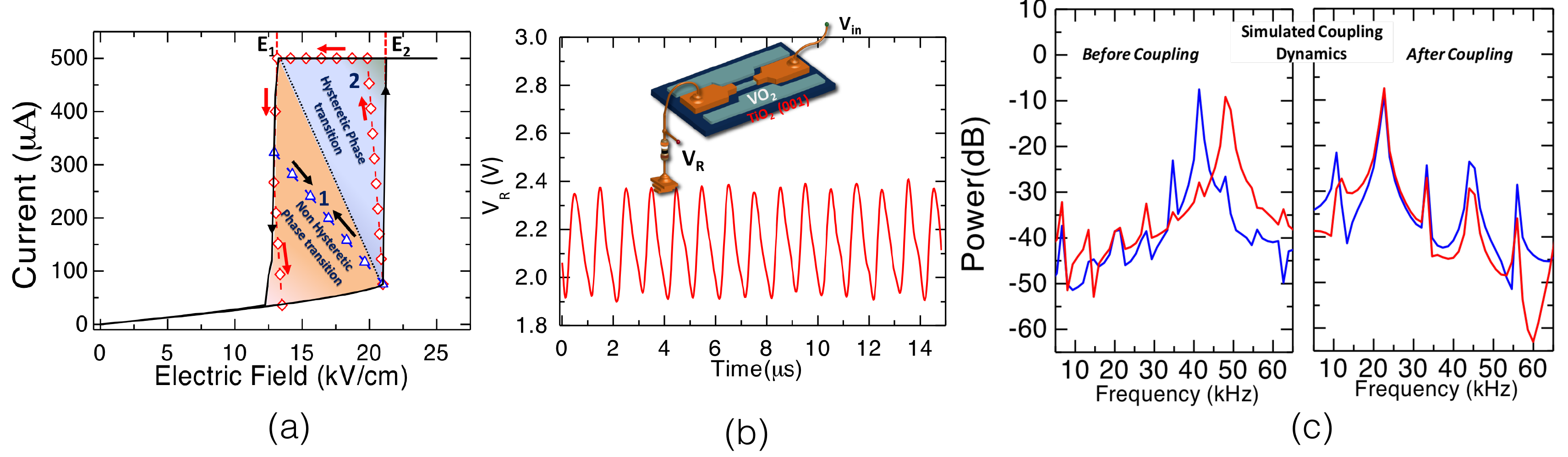}
\par\end{centering}
\caption{(a) Measured I-E characteristics of a VO2 device illustrating the hysteretic window (b) Measured characteristics show oscillations when biased with a pull down resistor, (c) Measurements and simulations illustrate frequency locking when two oscillators are coupled electrically. \label{fig:vo2_device}}
\end{figure}

Vanadium dioxide undergoes a first order metal-insulator phase transition marked by an abrupt change in conductivity up to five orders in magnitude. This abrupt change in conductivity which is also accompanied by changes in optical properties and can be triggered externally using thermal~\cite{Qazilbash1750}, optical~\cite{PhysRevLett.87.237401}, electronic~\cite{PhysRevB.77.235401} or strain~\cite{cite-key-nnano} stimuli can pave the way for novel electronic devices including steep slope switching devices, memory elements and ultra-fast optical switches. One of the characteristic features of the electrically driven first order metal-insulator transition in VO2 is its inherently hysteretic nature as shown in figure~\ref{fig:vo2_device}(a). In previous publications, we demonstrated an electric field driven non-hysteretic phase transition in VO2 and showed novel device functionalities like coupled oscillations that may enable  efficient implementation of novel, non-Boolean computing models. The electrically driven first-order phase transition in results in abrupt switching in conductivity but always comes at the cost of an intrinsic hysteresis, because the electrical field at which insulator to metal (IMT) transition occurs is always higher than that at which the metal to insulator transition happens (MIT). HIN our experiments, the VO2 is epitaxially grown on (001) TiO2 (-0.86\% compressive strain) using Molecular Beam Epitaxy (MBE), then patterned to form channels and followed by deposition of Au/Pd contacts to electrically access the VO2 channel (details of growth and fabrication are given the supplementary section). First, the device is electrically driven across the phase transition boundary with zero external series resistance. A current compliance is set to limit the current in the metallic state (to prevent excess joule heating resulting in permanent damage). The insulator to metal transition (IMT) and the reverse metal to insulator transition (MIT) occur at two critical fields, $E_2$ and $E_1$, respectively. The critical field $E_2$ is the field required to attain the Mott criteria and thereby trigger the formation of the metallic phase. Since transport in the insulating phase is dominated by hopping transport, we use the field dependent hopping conductance to understand its electrical properties.             .
The Negative Differential Resistance (NDR) region, where the VO2 is characterized by a conductivity intermediate between the metallic and insulating states  is referred to as the phase coexistence region. The net conductivity in this region is due to contributions from the metallic and insulating phase4. In-situ nano XRD characterization performed simultaneously with the transient waveform measurement confirms that the nature of the insulating phase is Monoclinic M1 and that of the metallic phase is Rutile which is expected as the films are -0.86\% compressively strained. When such a phase transition device is operated in the hystertic region, it breaks into spontaneous oscillations as shown in figure~\ref{fig:vo2_device}(b). This is a relaxation oscillator with piecewise linear dynamics as has been discussed in the previous section.

\subsection{Coupling IMT Oscillators: Electrical and Optical}
In our experiments and model development, we focus on electrical coupling using resistors and capacitors. The coupling dynamics has been discussed in the previous section and in more details in our earlier publications~\cite{parihar2015synchronization,parihar2014exploiting}. For example, two oscillators with slightly different native frequencies when coupled show frequency locking as observed in experiments and simulations~\ref{fig:vo2_mosfet}(c). A major challenge in scaling up chip-scale coupled oscillator circuits to solve practical computational problems lies in the difficulty in implementing an all-to-all oscillator coupling schemes. Recently, it has been proposed that cavity field coupled oscillators can overcome this challenge~\cite{PhysRevE.85.066203}. Lipson et. al.~\cite{PhysRevLett.109.233906} has demonstrated experimentally the synchronization of a pair of silicon nitride optomechanical oscillators (OMOs)  that are optically coupled through the radiation pressure field as opposed to mechanically coupled.  The researchers were able to turn the optical coupling on and off using a heating laser via thermooptic effect. The ability to manipulate the coupling strength using well known nanophotonics techniques such as colossal electro-optic or thermo-optic effects, increase their future potential to realize large scale on-chip nonlinear dynamical systems. This is an area of research which lies unexplored from a computational point of view, and success in creating complex systems with many to many connections will be key to achieving hardware platforms capable to truly delivering the promise of coupled dynamical systems as computational elements.

\section{Oscillator Networks and Applications}

\subsection{Image Data Processing and Analytics}
\begin{figure}[b]
\begin{centering}
\includegraphics[width=\textwidth]{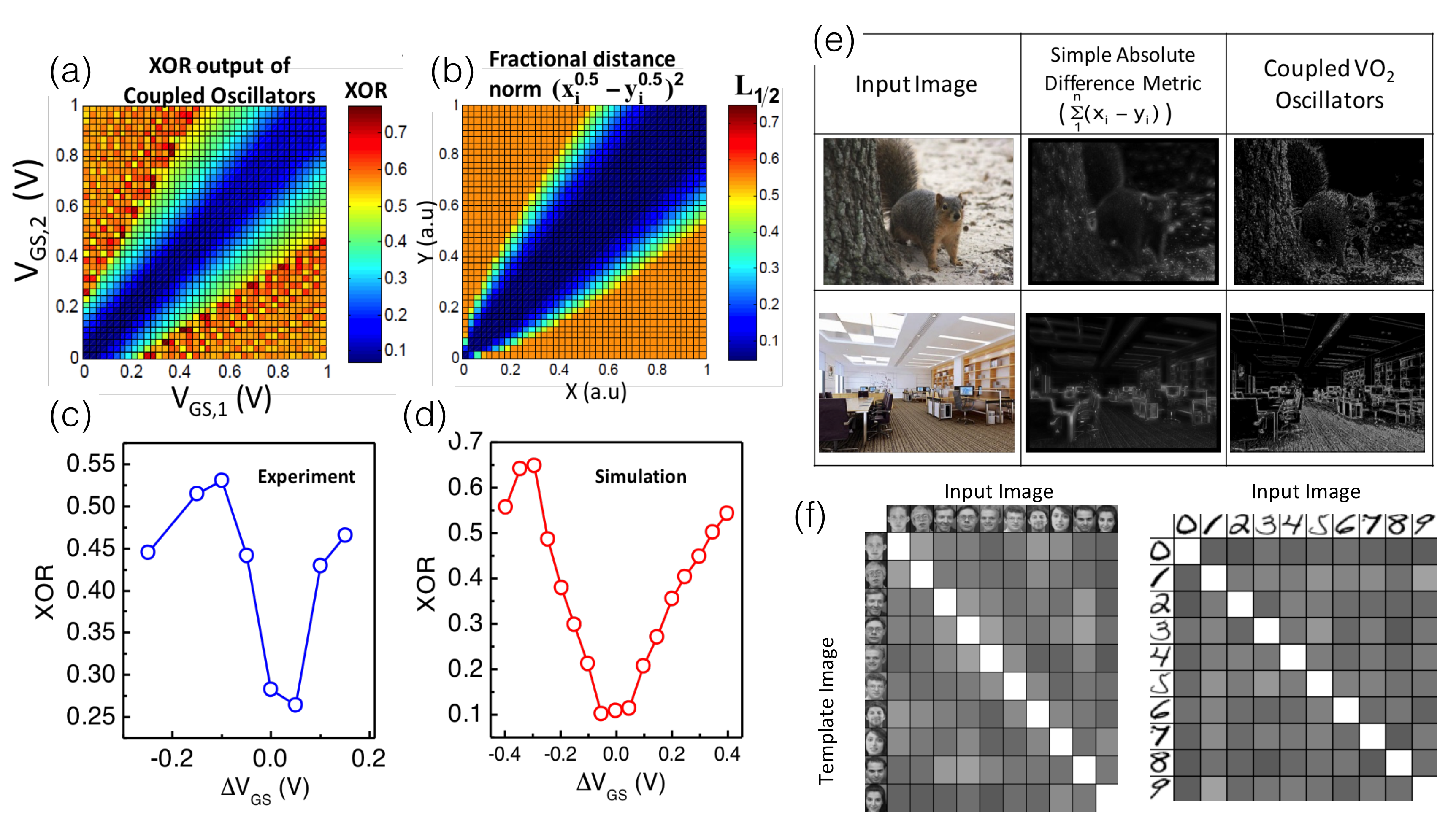}
\par\end{centering}
\caption{(a,b) XOR output of pairwise coupled oscillators show close similarity with fractional norm measure between two inputs (c,d) Experimental results and simulations reveal the capability of coupled oscillator phases to encode a measure of difference between two inputs ($\Delta V_{GS}$), (e) Saliency detection of images using coupled oscillator systems and a digital implementation, (f) a pairwise XOR measure is extended to illustrate pattern matching between an input and a template. A lighter color corresponds to a better match.}
\label{fig:iedm} 
\end{figure}

Arrays of such IMT oscillator based comparators can be used for template matching applications where element-wise comparisons suffice to decide a match. Figure~\ref{fig:iedm}(a,b) illustrate the XOR-ed output of the phases of two coupled oscillators configured as figure~\ref{fig:vo2_mosfet}. We observe from figure~\ref{fig:iedm}(c,d) a close match between experiments and simulations and it demonstrates that the XOR measure between the outputs of the two oscillators is a measure of distance between two inputs as given by $\Delta V_{GS} = V_{GS1}-V_{GS2}$.

We investigate the application of pairwise coupled oscillators for visual saliency approximations (detecting parts of the image that visually standout). Oscillator-based edge detection is performed using an array of pairwise oscillators to approximate the degree of dissimilarity between a given image pixel and its immediate neighbors. Different edges, vertical, horizontal, diagonal, are detected based on the selection of neighboring pixels for comparison. As this concept is expanded to include the comparison of pixels within a larger neighborhood (pixels surrounding reference pixel; a 3x3 neighborhood is used here), the output approximates the visual saliency. We also note that pairwise coupling of oscillators lead to the inherent calculation of fractional norms between templates and inputs, a task which is notoriously painstaking on a digitally abstracted computer. It is evident from figure~\ref{fig:iedm}(e) that the coupled oscillators show higher sensitivity to image contrast in comparison to a CMOS ASIC accelerator that uses a linear norm. 

We further demonstrate this by comparing images of faces and hand-written numbers as shown in figure~\ref{fig:iedm}(f). We first use the XOR measure for each pixel and calculate the number of pixels with XOR output below a threshold value . Figure~\ref{fig:iedm}(f) shows the results of comparing faces with a relaxation comparator, where the grey shade corresponds to the fraction of pixels with positive match, white being the highest. Such system followed by a winner-take-all (WTA), i.e. a threshold on the number of pixels that give a positive match, can be used to decide if the input image matches a stored template pattern. The value of  is chosen around 0.2 considering the minimum values of the XOR surface in the operating range of  values. The two thresholds described above depend on different factors. The threshold of the number of pixels for WTA would depend on the database and the error statistics required or estimated. On the other hand,  would be decided more by the nature of the XOR surface (Figure 10) and its minimum values.
A coupled VO2-MOSFET configuration cascaded with a XOR provides a way of measuring a form of fractional distance using FSK. Such associative networks can be used in more complex pattern matching and classification problems with potentially large benefits in energy efficiency.
The advantage of such oscillator based computing systems can be truly harnessed is they are miniaturized, made compact and integrated. We expect such systems to provide large improvements in energy-efficiency opening up possibilities in areas as varied as surveillance, consumer electronics and in-sensor processing. For comparison, a digital baseline design is designed and simulated. All digital circuits are implemented with 11nm node transistor models. We observe that the coupled oscillators provide a power reduction of ~20X over CMOS reflecting the advantage of `let physics do the computing' approach and potentially removing the Boolean bottleneck. For further details interested readers are pointed to the previous work by the authors~\cite{shukla2014pairwise}.

\subsection{Complex Global Connections and Possibilities for Computation}

\begin{figure}
\begin{centering}
\includegraphics{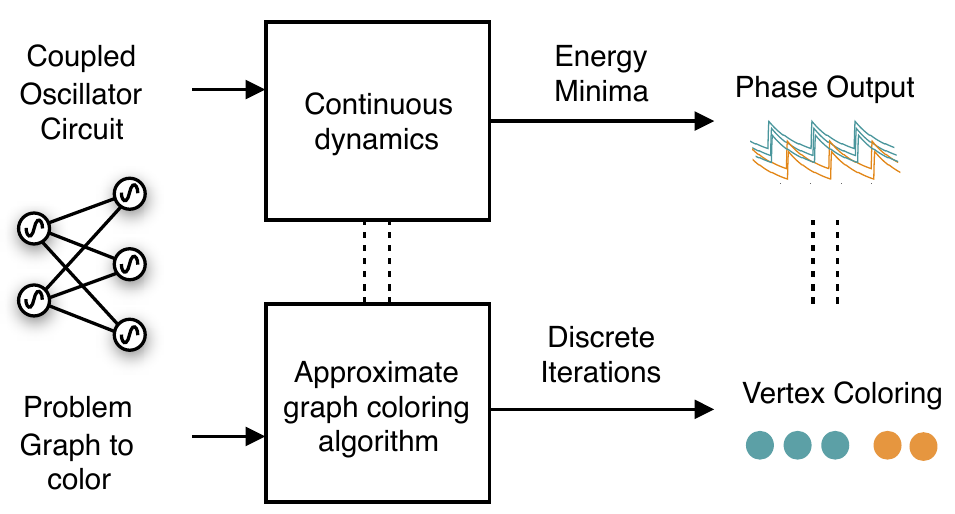}
\par\end{centering}
\caption{Overview of the proposed graph coloring system\label{fig:graph_coloring}}
\end{figure}

The full computational power of such coupled dynamical systems can
be further harnessed using complex global interactions among oscillators instead
of just pairwise interactions. Such coupled oscillators provide a Hopfield type networks, but with piecewise linear dynamics and hysteretic
switching of oscillators. When connected in complex
networks with both global and local connectivity, instead of pairwise coupling, these networks of oscillators can be much more powerful and
can possibly compute approximate solutions of many hard optimization
problems. In one implementation of globally coupled D-R oscillators
(with a series resistance), we have recently demonstrated using theory and
experimental implementations that such networks are capable of approximating
the solution of NP-hard minimum graph coloring problems. When $n$
identical D-R oscillators are coupled using identical capacitors,
such network settles to a steady state wherein the relative phases
of the oscillators get ordered in a way that corresponds to the solution
of graph coloring. The time evolution of the piecewise linear dynamical
system of such coupled D-R oscillators inherits the properties of
and hence mimics approximate graph coloring algorithms because of
the construction of such networks. Basic functioning of such a system
is shown in figure \ref{fig:graph_coloring}.

\section{Concluding Remarks}
In this article, we have demonstrated that the dynamics that evolve from the complex interactions among oscillators can be a powerful computing paradigm. However, this requires innovations in fabrication of compact and coupled networks of oscillators; and the current implementation using phase transition devices is one promising candidate. Even with advances in post-silicon devices and technology, the true potential of integrated dynamical systems can only be harnessed when controllable and programmable coupling can be realized and the phase and frequency dynamics carefully measured and read out. A plethora of challenges remain; however the opportunities offered by dynamical systems to make a significant impact in a post-CMOS world in undeniable.

\bibliographystyle{unsrt}
\bibliography{refs}

\begin{thebibliography}{10}

\bibitem{garey1979computers}
Michael~R. Garey and David~S. Johnson.
\newblock {\em Computers and {Intractability}: {A} {Guide} to the {Theory} of
  {NP}-{Completeness}}.
\newblock W. H. Freeman \& Co., New York, NY, USA, 1979.

\bibitem{wolfram1986theoryand}
Stephen Wolfram, editor.
\newblock {\em Theory and applications of cellular automata: including selected
  papers, 1983-1986}.
\newblock Number v. 1 in Advanced series on complex systems. World Scientific,
  Singapore, 1986.

\bibitem{shor1997polynomialtime}
P.~Shor.
\newblock Polynomial-{Time} {Algorithms} for {Prime} {Factorization} and
  {Discrete} {Logarithms} on a {Quantum} {Computer}.
\newblock {\em SIAM Journal on Computing}, 26(5):1484--1509, October 1997.

\bibitem{lucas2014isingformulations}
Andrew Lucas.
\newblock Ising formulations of many {NP} problems.
\newblock {\em Interdisciplinary Physics}, 2:5, 2014.

\bibitem{wang2013coherent}
Zhe Wang, Alireza Marandi, Kai Wen, Robert~L. Byer, and Yoshihisa Yamamoto.
\newblock Coherent {Ising} machine based on degenerate optical parametric
  oscillators.
\newblock {\em Physical Review A}, 88(6):063853, December 2013.

\bibitem{hopfield1985textquotedblleftneuraltextquotedblright}
J.~J. Hopfield and D.~W. Tank.
\newblock {\textquotedblleft}{Neural}{\textquotedblright} computation of
  decisions in optimization problems.
\newblock {\em Biological Cybernetics}, 52(3):141--152, July 1985.

\bibitem{mostafa2015aneventbased}
Hesham Mostafa, Lorenz~K. M{\"u}ller, and Giacomo Indiveri.
\newblock An event-based architecture for solving constraint satisfaction
  problems.
\newblock {\em Nature Communications}, 6:8941, December 2015.

\bibitem{traversa2015memcomputing}
F.~L. Traversa, C.~Ramella, F.~Bonani, and M.~Di~Ventra.
\newblock Memcomputing {NP}-complete problems in polynomial time using
  polynomial resources and collective states.
\newblock {\em Science Advances}, 1(6):e1500031--e1500031, July 2015.

\bibitem{chua1988cellular2}
L.~O. Chua and L.~Yang.
\newblock Cellular neural networks: applications.
\newblock {\em IEEE Transactions on Circuits and Systems}, 35(10):1273--1290,
  October 1988.

\bibitem{chua1988cellular}
L.~O. Chua and L.~Yang.
\newblock Cellular neural networks: theory.
\newblock {\em IEEE Transactions on Circuits and Systems}, 35(10):1257--1272,
  October 1988.

\bibitem{toffoli1984cellular}
Tommaso Toffoli.
\newblock Cellular automata as an alternative to (rather than an approximation
  of) differential equations in modeling physics.
\newblock {\em Physica D: Nonlinear Phenomena}, 10(1):117--127, January 1984.

\bibitem{siegelmann1995computation}
Hava~T. Siegelmann.
\newblock Computation {Beyond} the {Turing} {Limit}.
\newblock {\em Science}, 268(5210):545--548, April 1995.

\bibitem{vergis1986thecomplexity}
Anastasios Vergis, Kenneth Steiglitz, and Bradley Dickinson.
\newblock The complexity of analog computation.
\newblock {\em Mathematics and Computers in Simulation}, 28(2):91--113, April
  1986.

\bibitem{imada1998metalinsulator}
Masatoshi Imada, Atsushi Fujimori, and Yoshinori Tokura.
\newblock Metal-insulator transitions.
\newblock {\em Reviews of Modern Physics}, 70(4):1039--1263, October 1998.

\bibitem{shukla2014synchronized}
Nikhil Shukla, Abhinav Parihar, Eugene Freeman, Hanjong Paik, Greg Stone,
  Vijaykrishnan Narayanan, Haidan Wen, Zhonghou Cai, Venkatraman Gopalan, Roman
  Engel-Herbert, Darrell~G. Schlom, Arijit Raychowdhury, and Suman Datta.
\newblock Synchronized charge oscillations in correlated electron systems.
\newblock {\em Scientific Reports}, 4, May 2014.

\bibitem{shukla2014pairwise}
N.~Shukla, A.~Parihar, M.~Cotter, M.~Barth, X.~Li, N.~Chandramoorthy, H.~Paik,
  D.G. Schlom, V.~Narayanan, A.~Raychowdhury, and S.~Datta.
\newblock Pairwise coupled hybrid vanadium dioxide-{MOSFET} ({HVFET})
  oscillators for non-boolean associative computing.
\newblock In {\em Electron {Devices} {Meeting} ({IEDM}), 2014 {IEEE}
  {International}}, pages 28.7.1--28.7.4, December 2014.

\bibitem{strogatz2000fromkuramoto}
Steven~H. Strogatz.
\newblock From {Kuramoto} to {Crawford}: exploring the onset of synchronization
  in populations of coupled oscillators.
\newblock {\em Physica D: Nonlinear Phenomena}, 143(1):1--20, 2000.
\newblock read.

\bibitem{vanderpol1934thenonlinear}
Balth Van Der~Pol.
\newblock The {Nonlinear} {Theory} of {Electric} {Oscillations}.
\newblock {\em Proceedings of the Institute of Radio Engineers},
  22(9):1051--1086, 1934.

\bibitem{rand1980bifurcation}
R.~H. Rand and P.~J. Holmes.
\newblock Bifurcation of periodic motions in two weakly coupled van der {Pol}
  oscillators.
\newblock {\em International Journal of Non-Linear Mechanics},
  15(4{\textendash}5):387--399, 1980.

\bibitem{storti1982dynamics}
D.~W. Storti and R.~H. Rand.
\newblock Dynamics of two strongly coupled van der pol oscillators.
\newblock {\em International Journal of Non-Linear Mechanics}, 17(3):143--152,
  1982.

\bibitem{kopell1995antiphase}
Nancy Kopell and David Somers.
\newblock Anti-phase solutions in relaxation oscillators coupled through
  excitatory interactions.
\newblock {\em Journal of Mathematical Biology}, 33(3):261--280, December 1995.

\bibitem{hirano2003existence}
Norimichi Hirano and S{\l }awomir Rybicki.
\newblock Existence of limit cycles for coupled van der {Pol} equations.
\newblock {\em Journal of Differential Equations}, 195(1):194--209, November
  2003.

\bibitem{7120163}
K.~Yogendra, D.~Fan, and K.~Roy.
\newblock Coupled spin torque nano oscillators for low power neural
  computation.
\newblock {\em IEEE Transactions on Magnetics}, 51(10):1--9, Oct 2015.

\bibitem{cite-key-kroy}
Abhronil Sengupta, Priyadarshini Panda, Parami Wijesinghe, Yusung Kim, and
  Kaushik Roy.
\newblock Magnetic tunnel junction mimics stochastic cortical spiking neurons.
\newblock {\em Scientific Reports}, 6:30039 EP --, 07 2016.

\bibitem{parihar2015synchronization}
Abhinav Parihar, Nikhil Shukla, Suman Datta, and Arijit Raychowdhury.
\newblock Synchronization of pairwise-coupled, identical, relaxation
  oscillators based on metal-insulator phase transition devices: {A} model
  study.
\newblock {\em Journal of Applied Physics}, 117(5):054902, February 2015.

\bibitem{datta2014neuroinspired}
Suman Datta, Nikhil Shukla, Matthew Cotter, Abhinav Parihar, and Arijit
  Raychowdhury.
\newblock Neuro {Inspired} {Computing} with {Coupled} {Relaxation}
  {Oscillators}.
\newblock In {\em Proceedings of the {The} 51st {Annual} {Design} {Automation}
  {Conference} on {Design} {Automation} {Conference}}, {DAC} '14, pages
  74:1--74:6, New York, NY, USA, 2014. ACM.

\bibitem{parihar2014exploiting}
A.~Parihar, N.~Shukla, S.~Datta, and A.~Raychowdhury.
\newblock Exploiting {Synchronization} {Properties} of {Correlated} {Electron}
  {Devices} in a {Non}-{Boolean} {Computing} {Fabric} for {Template}
  {Matching}.
\newblock {\em IEEE Journal on Emerging and Selected Topics in Circuits and
  Systems}, PP(99):1--10, 2014.

\bibitem{Qazilbash1750}
M.~M. Qazilbash, M.~Brehm, Byung-Gyu Chae, P.-C. Ho, G.~O. Andreev, Bong-Jun
  Kim, Sun~Jin Yun, A.~V. Balatsky, M.~B. Maple, F.~Keilmann, Hyun-Tak Kim, and
  D.~N. Basov.
\newblock Mott transition in vo2 revealed by infrared spectroscopy and
  nano-imaging.
\newblock {\em Science}, 318(5857):1750--1753, 2007.

\bibitem{PhysRevLett.87.237401}
A.~Cavalleri, Cs. T\'oth, C.~W. Siders, J.~A. Squier, F.~R\'aksi, P.~Forget,
  and J.~C. Kieffer.
\newblock Femtosecond structural dynamics in ${\mathrm{vo}}_{2}$ during an
  ultrafast solid-solid phase transition.
\newblock {\em Phys. Rev. Lett.}, 87:237401, Nov 2001.

\bibitem{PhysRevB.77.235401}
Bong-Jun Kim, Yong~Wook Lee, Sungyeoul Choi, Jung-Wook Lim, Sun~Jin Yun,
  Hyun-Tak Kim, Tae-Ju Shin, and Hwa-Sick Yun.
\newblock Micrometer x-ray diffraction study of ${\text{vo}}_{2}$ films:
  Separation between metal-insulator transition and structural phase
  transition.
\newblock {\em Phys. Rev. B}, 77:235401, Jun 2008.

\bibitem{cite-key-nnano}
CaoJ., ErtekinE., SrinivasanV., FanW., HuangS., ZhengH., YimJ.~W. L., KhanalD.
  R., OgletreeD. F., GrossmanJ. C., and WuJ.
\newblock Strain engineering and one-dimensional organization of
  metal-insulator domains in single-crystal vanadium dioxide beams.
\newblock {\em Nat Nano}, 4(11):732--737, 11 2009.

\bibitem{PhysRevE.85.066203}
C.~A. Holmes, C.~P. Meaney, and G.~J. Milburn.
\newblock Synchronization of many nanomechanical resonators coupled via a
  common cavity field.
\newblock {\em Phys. Rev. E}, 85:066203, Jun 2012.

\bibitem{PhysRevLett.109.233906}
Mian Zhang, Gustavo~S. Wiederhecker, Sasikanth Manipatruni, Arthur Barnard,
  Paul McEuen, and Michal Lipson.
\newblock Synchronization of micromechanical oscillators using light.
\newblock {\em Phys. Rev. Lett.}, 109:233906, Dec 2012.

\end{thebibliography}

\end{document}